\documentclass{article}
\usepackage{spconf,amsmath,graphicx,hyperref}

\usepackage{booktabs} 
\usepackage{amsfonts}

\title{PerformSinger: Multimodal Singing Voice Synthesis Leveraging\\ Synchronized Lip Cues from Singing Performance Videos}

\name{Ke Gu$^{1}$, Zhicong Wu$^{1}$, Peng Bai$^{1}$, Sitong Qiao$^{2}$, Zhiqi Jiang$^{2}$, Junchen Lu$^{3}$, Xiaodong Shi$^{1}$, Xinyuan Qian$^{2}$}
\address{$^{1}$Xiamen University \\
        $^{2}$University of Science and Technology Beijing \\
         $^{3}$National University of Singapore\\
         }

\begin{document}

\ninept
\maketitle
\begin{abstract}

Existing singing voice synthesis (SVS) models largely rely on fine-grained, phoneme-level durations, which limits their practical application. These methods overlook the complementary role of visual information in duration prediction.
To address these issues,  we propose PerformSinger, a pioneering multimodal SVS framework, which incorporates lip cues from video as a visual modality, enabling high-quality ``duration-free" singing voice synthesis.
PerformSinger comprises parallel multi-branch multimodal encoders, a feature fusion module, a duration and variational prediction network, a mel-spectrogram decoder and a vocoder. 
The fusion module, composed of adapter and fusion blocks, employs a progressive fusion strategy within an aligned semantic space to produce high-quality multimodal feature representations, thereby enabling accurate duration prediction and high-fidelity audio synthesis. 
To facilitate the research, we design, collect and annotate a novel SVS dataset involving synchronized video streams and precise phoneme-level manual annotations. 
Extensive experiments demonstrate the state-of-the-art performance of our proposal in both subjective and objective evaluations. The code and dataset will be publicly available.

\end{abstract}

\section{Introduction}

Recent advances in Singing Voice Synthesis (SVS) have enabled the generation of high-fidelity and expressive voices. Representative models such as HiFiSinger~\cite{chen2020hifisinger}, DiffSinger~\cite{liu2022diffsinger}, and VISinger2~\cite{zhang2022visinger} have significantly advanced synthesis quality by introducing high-fidelity vocoders, diffusion mechanisms, and differentiable synthesizers. Meanwhile, various studies have investigated approaches to enable more flexible and fine-grained control of synthesized singing voices. StyleSinger~\cite{zhang2024stylesinger} captures style features from out-of-domain references; Prompt-Singer~\cite{wang2024prompt} uses natural-language prompts; and ExpressiveSinger~\cite{dai2024expressivesinger} explicitly models performance timing and dynamics from sheet scores. 

Despite notable advancements in audio quality, emotion, and controllability, previous SVS methods still rely heavily on fine-grained, phoneme-level duration inputs. This dependency not only limits their usability during inference but also raises the entry barrier for users. Furthermore, prior SVS research has primarily focused on text and reference audio, overlooking the complementary role of visual information~\cite{lu2022visualtts}.
Evidence from related studies further supports this potential: MM-ALT improves lyric transcription by jointly leveraging lip video and singing audio~\cite{gu2022mm,gu2024automatic}, URSing utilizes visual features for singing voice separation~\cite{li2021audiovisual}, and A cappella exploits facial keypoints to enhance separation~\cite{montesinos2021cappella}. Although these works do not target SVS directly, they demonstrate that lip movements in singing videos are strongly correlated with textual content and can provide additional cues for accurate duration prediction.

\begin{figure}[t]
\centering
\includegraphics[width=\columnwidth]{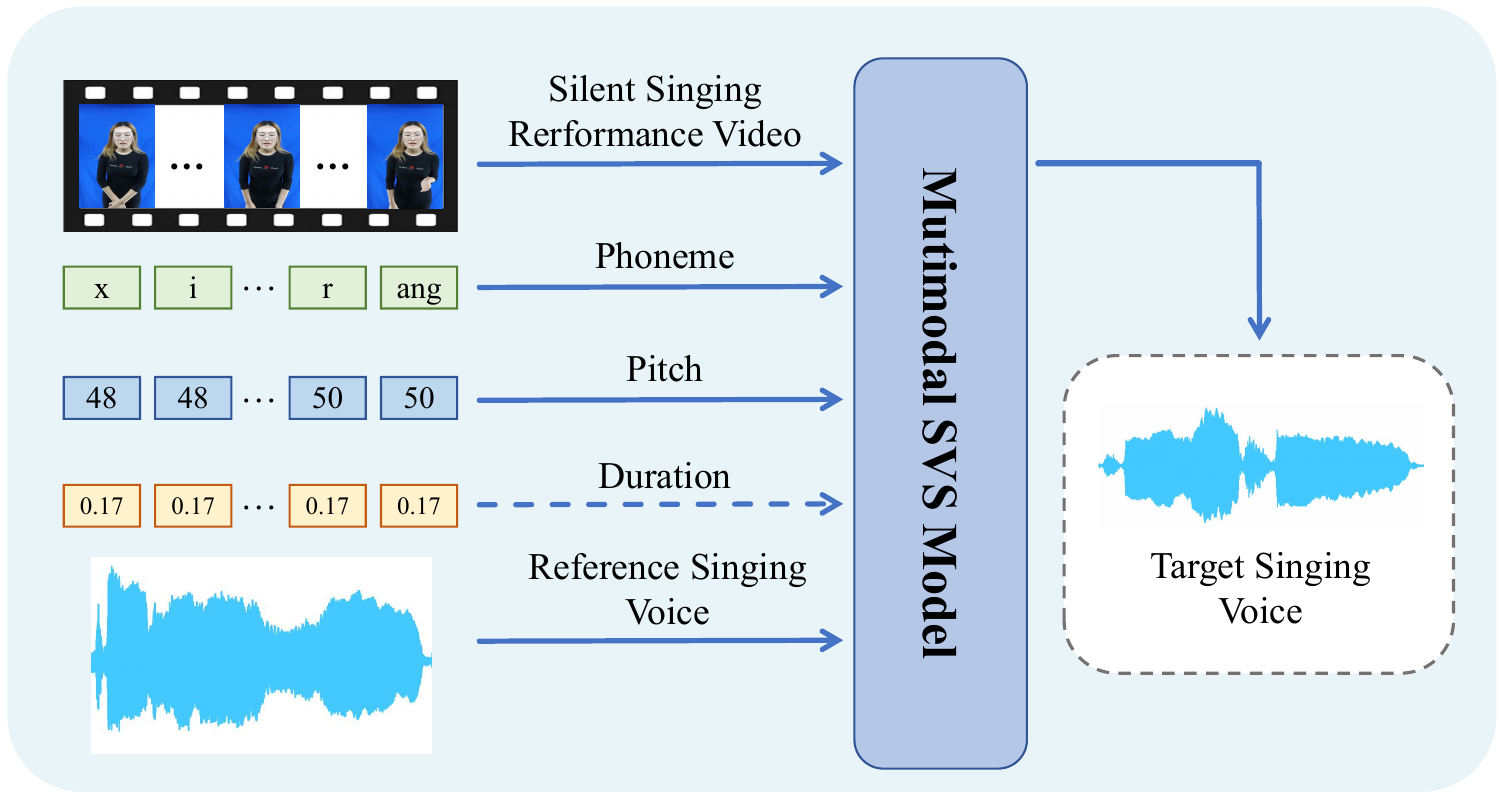}
\caption{Overview of our proposed multimodal singing voice synthesis task, where duration is not required as input (dashed lines).}
\label{fig:frame}
\end{figure}

Consequently, we extend the SVS task by introducing singing-performance video inputs, leveraging lip-movement sequences to eliminate the need for explicit duration inputs,  as illustrated in Figure \ref{fig:frame}. To support this, we construct VisualSinger, a multimodal SVS dataset with utterance-level audio–video pairs and precise phoneme and pitch annotations.

Building on this dataset, we propose PerformSinger, a novel multimodal SVS framework that encodes audio, text, and video in parallel, incorporating visual lip cues to achieve duration-free synthesis. Our main contributions are summarized as follows:
\begin{itemize}
\item We propose PerformSinger, a multimodal SVS model that exploits lip cues from singing video  to achieve duration-free synthesis.
\item We construct and release VisualSinger, the first audio-visual SVS dataset, which includes precise utterance segmentation and comprehensive phoneme and pitch annotations to support future multimodal research.
\item We design an effective and novel multimodal framework that fuses multimodal features within an aligned semantic space, significantly enhancing both duration prediction and synthesis quality.
\end{itemize}
Experiments on our proposed dataset show that the proposed PerformSinger surpasses other models in both subjective and objective evaluations, demonstrating the effectiveness of our framework. 

\section{VisualSinger}
This section introduces VisualSinger, a SVS dataset that comprises over three hours of synchronized audio-visual singing streams with precise phoneme-level annotations, with the pipeline and statistics.

\subsection{Data Sources}

We initially surveyed public datasets from related audio-visual singing tasks, including audio-visual singing-voice separation and lyrics-and-music transcription. Of these, only the URSing\cite{li2021audiovisual} dataset aligns with our requirements, providing song-level performances with isolated vocals and matched high-fidelity audio-visual data. Despite URSing lacking the aligned lyrics, utterance-level segmentation, and phoneme-level pitch labels essential for SVS, we still utilized it as a foundational data source for further processing.
To expand the data, we additionally collected singing videos from the web. 
The results were then manually filtered to retain only authentic videos (i.e., not lip-synced) where singer directly faces the camera, thereby ensuring the authenticity of visual cues. Consequently, our final dataset is primarily composed of the Chinese subset of URSing and the Chinese singing videos collected from the web.

\subsection{Dataset Construction and Annotation}

After collecting the singing videos, we processed them into utterance-level audio-video-text triplets through a structured pipeline. We first separated vocals from accompaniment using MDX-KimVocal2, then isolated lead vocals with VR 5HP Karaoke\footnote{https://github.com/nomadkaraoke/python-audio-separator}. Lyrics were identified by video titles and musical content. Utterances were segmented at silence boundaries (5–15s) using silero-vad\footnote{https://github.com/snakers4/silero-vad} and manually refined in Audacity to preserve lyrical completeness. The final timestamps were used to synchronously extract audio and video clips via FFmpeg. Chinese lyrics were converted to phonemes by pypinyin.

Because singers often apply key shifts, improvisations, and embellishments, we first obtain word and phoneme boundaries before extracting note pitches from the actual singing. We trained Montreal Forced Aligner (MFA) \cite{mcauliffe2017montreal} on our audio–text corpus to generate initial and coarse alignments. Given MFA's low accuracy on Chinese singing, five musically trained annotators manually review the TextGrid files in Praat, correcting mismatched words/phonemes, aligning boundaries by hearing, and marking silences and breaths. Using these fine-grained boundaries, F$_0$ is extracted and averaged with Parselmouth~\cite{jadoul2018introducing} for coarse pitch estimation, then refined via the pre-trained ROSVOT model~\cite{li2024robust}. 
Among these steps, manual TextGrid annotation is the most labor-intensive, requiring 40–45 human-hours per hour of audio for boundary and pitch confirmation.

\subsection{Statistics}

The VisualSinger dataset involves 9 singers performing 69 songs. These songs are segmented into 1,248 utterances, with a total duration of 11,080 seconds (approximately 3.07 hours). In terms of phonemes, our dataset contains 21 initials and 35 finals, covering all possible initials. Only the rare final “ueng” is missing (its frequency in daily vocabulary is less than 0.01\%)~\cite{zhang2022m4singer}.Among the initials, \textit{d} is most frequent and \textit{p} least; for the finals, \textit{i} appears most often, while \textit{vn} is rarest.
In terms of musical notes, the pitch ranges from 36 to 79. The majority are distributed between 50 and 62, concentrated in the midrange corresponding to typical male and female vocal ranges.

\section{Methodology}

\begin{figure*}[t]
\centering
\includegraphics[width=0.83\textwidth]{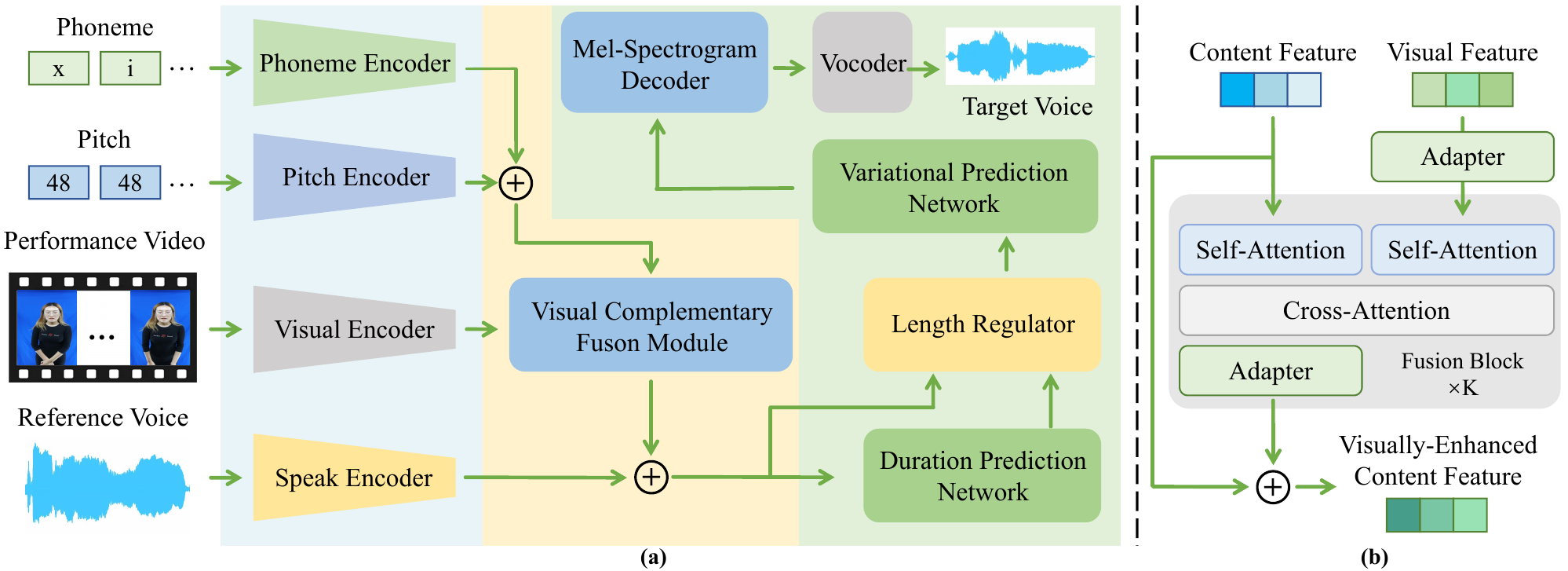}
\caption{PerformSinger model architecture: (a) overall structure; (b) Visual Complementary Fusion Module (VCFM) details.}
\label{fig:network}
\end{figure*}

\subsection{Problem Formulation}

We formulate the multimodal, duration-free singing voice synthesis task.
Given a silent video $V = (v_1, \dots, v_m)$, a reference audio clip $A_\mathrm{ref} = (a_1, \dots, a_p)$, a lyric sequence $T = (t_1, \dots, t_n)$, and the aligned pitch sequence $P = (p_1, \dots, p_n)$, the objective is to generate target singing audio $A_\mathrm{tgt} = (a_1, \dots, a_q)$ that synchronizes with the video, matches the lyrics and pitch contour, and preserves the reference timbre.
Compared to traditional SVS, our task incorporates the visual stream while removing the explicit duration sequence.

\subsection{Framework Overview}

As shown in Figure~\ref{fig:network}(a), our framework first employs modality--specific encoders to independently process the text, pitch, reference audio, and video streams. The extracted multimodal features are then integrated by a visual complementary fusion module, after which the fused representation is routed through a series of predictor networks and a decoder to generate a mel‑spectrogram that is ultimately converted into an audio waveform by a vocoder.

\subsection{Parallel Multi-branch Multimodal Encoders}

We adopt a FastSpeech2–style \cite{ren2020fastspeech} phoneme encoder $\mathrm{Enc}_{\mathrm{T}}$ combining self-attention and 1-D convolutions for global–local dependency modeling; The pitch encoder $\mathrm{Enc}_{\mathrm{P}}$ is a learnable embedding layer mapping. The entire process can be formulated as:

\begin{gather}
TF = \mathrm{Enc}_{\mathrm{T}}(T),\;
PF = \mathrm{Enc}_{\mathrm{P}}(P),\\
CF = TF + PF,
\end{gather}

where $T$ and $P$ denote the input phoneme sequence and pitch sequence, respectively, and $TF$, $PF$, and $CF$ represent the encoded textual feature, pitch feature, and their summed content feature.

For the reference audio, we adopt the d‑vector speaker encoder $\mathrm{Enc}_{\mathrm{S}}$ proposed by~\cite{wan2018generalized}, which extracts a robust timbre vector to provide the speaker condition for singing voice synthesis.
For the video input, we crop each frame's mouth region to form a lip image sequence, then extract visual cues using a widely used pretrained lip‑reading encoder~\cite{wu2019time}, composed of a 3D convolution block and an 18‑layer ResNet, which has proven effective in audio–visual speech separation and synthesis~\cite{afouras2018conversation,gu2020multi,lu2022visualtts}. The audio and visual encoding processes can be jointly formulated as:

\begin{gather}
SF = \mathrm{Enc}_{\mathrm{S}}(A_{\mathrm{ref}}), \quad
VF = \mathrm{Enc}_{\mathrm{V}}(V),
\end{gather}
where $A_{\mathrm{ref}}$ and $V$ denote the reference speech and lip image sequence, respectively, and $SF$ and $VF$ are the corresponding speaker and lip feature representation produced by the encoders.

\subsection{Visual Complementary Fusion Module}

Lip shape features provide rich viseme cues for phoneme timing and duration estimation. To leverage this visual information, we design the Visual Complementary Fusion Module (VCFM). As shown in Figure~\ref{fig:network}(b), we first align cross-modal feature distributions using an adapter to enhance representation consistency. The adapter uses an up-projection–nonlinearity–down-projection structure:
\begin{equation}
VF^{\prime} = L_{1}\bigl(\mathrm{GELU}(L_{2}(VF))\bigr),
\end{equation}
where $L_{2}:\mathbb{R}^{H}\to\mathbb{R}^{2H}$ expands the channel dimension from $H$ to $2H$ and $L_{1}:\mathbb{R}^{2H}\to\mathbb{R}^{H}$ projects back to $H$. 

Next, we inject visual features into content features $CF$ via a residual fusion block to integrate visual cues into phoneme modeling. To reduce cross-modal gaps and stabilize fusion, we employ a progressive strategy stacking $K$ fusion blocks. Each block first applies self-attention to $CF^{(i-1)}$ and $VF'$, then uses cross-attention with queries from $\mathrm{SA}(CF^{(i-1)})$ and keys/values from $\mathrm{SA}(VF')$, followed by LayerNorm and the adapter, with a residual connection. This process can be described by the following equations:
\begin{equation}
Z^{(i)} = \mathrm{LN}\bigl(\mathrm{CA}(\mathrm{SA}(VF'),\,\mathrm{SA}(CF^{(i-1)}))\bigr),
\end{equation}
\begin{equation}
CF^{(i)} = CF^{(i-1)} + \mathrm{Adapter}(Z^{(i)}),
\end{equation}
where $i\in[1,K]$, $CF^{(0)}=CF$, and $VF'$ remains constant across blocks. $Z^{(i)}$ is the intermediate feature inside the module obtained after attention-based fusion. LN, SA and CA denote LayerNorm, Self-Attention and Cross-Attention, respectively. 
Through this design, we achieve effective integration of visual and content modalities, enhancing the model’s use of visual cues.

\begin{table*}[t]
\centering
\setlength{\tabcolsep}{4pt}
\caption{Objective and subjective evaluation results on our dataset under setting 1. (Bold indicates the best results, and underline indicates the second-best results). Subjective metrics are reported with 95\% confidence intervals.}
\label{tab:main}
\scalebox{0.95}{
\begin{tabular}{c|ccc cc|cccc}
\toprule
\textbf{Model} & \textbf{MCD$\downarrow$} & \textbf{FFE$\downarrow$} & \textbf{COS$\uparrow$} & \textbf{LSE-C$\uparrow$} & \textbf{LSE-D$\downarrow$} & \textbf{MOS-Q$\uparrow$} & \textbf{MOS‑N$\uparrow$} & \textbf{MOS‑S$\uparrow$} & \textbf{MOS‑M$\uparrow$} \\
\midrule
GT                & 0.1122 & 0.0002 & 0.9994 & 3.8862 & 8.0946 &     --    &    --     &    --     &    --     \\
\cmidrule(lr){1-10}
HPMDubbing        & 9.4733 & 0.7698 & 0.5194 & 0.8945 & 11.7521 & 1.45 $\pm$ 0.51 & 1.63 $\pm$ 0.62 & 1.41 $\pm$ 0.57 & 1.56 $\pm$ 0.55 \\
HPMDubbing-P & 9.4677 & 0.7629 & 0.5170 & 0.9102 & 11.7158 & 1.46 $\pm$ 0.55 & 1.82 $\pm$ 0.53 & 1.59 $\pm$ 0.58 & 1.73 $\pm$ 0.49 \\
StyleSinger       & \underline{3.3580} & \underline{0.4494} & \underline{0.9141} & \underline{1.3057} & \underline{10.4129} & \underline{3.54 $\pm$ 0.37} & \underline{3.79 $\pm$ 0.35} & \underline{3.62 $\pm$ 0.47} & \underline{3.73 $\pm$ 0.48} \\
\cmidrule(lr){1-10}
Our               & \textbf{3.1125} & \textbf{0.3921} & \textbf{0.9206} & \textbf{1.4270} & \textbf{10.2782} & \textbf{3.71 $\pm$ 0.34} & \textbf{3.88 $\pm$ 0.52} & \textbf{3.75 $\pm$ 0.36} & \textbf{3.82 $\pm$ 0.42} \\
\bottomrule
\end{tabular}}
\end{table*}

\subsection{Task‑Specific Prediction Network and Decoder}

This part specifically comprises:

\textbf{Duration Prediction Network.} The duration prediction network adopts a FastSpeech2-style architecture based on stacked 1D convolutions, taking content features fused with visual information as input to predict each phoneme's mel-frame count. 
A Length Regulator then expands the content features along the time axis according to the predicted frame count, ensuring precise alignment between the content representation and the target audio.

\textbf{Variational Prediction Network.} With the frame-level content representation enriched by visual features, we proceed to predict frame-wise pitch and style feature.
Given the complex pitch variations in singing voices, we employ a diffusion-based pitch prediction network following\cite{he2023rmssinger} to predict F0 and Unvoiced/Voiced labels.
To extract fine-grained style features from the reference audio, we introduce a style extraction network that uses residual vector quantization (RVQ) as a bottleneck to disentangle style and applies attention to align style features with content. See \cite{zhang2024stylesinger} for details.

\textbf{Mel‑Spectrogram Decoder.} Due to the complexity of singing's fine-grained dynamics, we adopt a diffusion-based mel-spectrogram decoder following DiffSinger, which generates the target spectrogram via iterative denoising conditioned on content features.

\subsection{Loss Function}

To optimize the synthesis quality, we design a composite loss function $\mathcal{L_{\text{Total}}}$ that integrates three four metrics including Mel‑spectrogram Reconstruction Loss ($\mathcal{L_{\text{R}}}$), Duration Loss ($\mathcal{L_{\text{D}}}$), Pitch Loss ($\mathcal{L_{\text{P}}}$) and Quantization Loss ($\mathcal{L_{\text{C}}}$).
Formally, the composite loss is defined as:
\begin{equation}
    \mathcal{L_{\text{Total}}} = \lambda_{R}\mathcal{L_{\text{R}}} 
    +\lambda_{D}\mathcal{L_{\text{D}}}
    +\lambda_{P}\mathcal{L_{\text{P}}}
    +\lambda_{C}\mathcal{L_{\text{C}}},
\end{equation}
where the $\lambda_{R}$, $\lambda_{D}$, $\lambda_{P}$ and $\lambda_{C}$ are hyperparameters that control the relative importance of each loss component.

\section{Experimental Setup}

\subsection{Dataset and Comparison Models}

Excluding two speakers (38 utterances) for out-of-domain testing, the rest seven speakers’ data is split 8:1:1 for training, validation, and test, yielding 967, 121, and 122 utterances.

To evaluate our proposed method, we select the following representative models as baselines: HPMDubbing~\cite{cong2023learning} is a visual dubbing model that produces lip-synchronized speech. We adapt it to the SVS task by augmenting it with phoneme-level pitch as input, denoting it as HPMDubbing-P.
StyleSinger~\cite{zhang2024stylesinger} is a representative reference audio-based SVS model. We configure it in a duration-free setting to enable fair evaluation under the same task conditions.

\subsection{Implementation Details}
We convert waveforms to 80-bin mel-spectrograms (48 kHz, 1024 FFT, 256 hop). The model features a 256-dim hidden space, a four-layer Transformer phoneme encoder, and 20-layer residual decoder, trained with Adam~\cite{kingma2014adam} ($\beta_1=0.9$, $\beta_2=0.98$) on two NVIDIA A40 GPUs; HiFi-GAN~\cite{kong2020hifi} serves as vocoder. A two-stage hierarchical training strategy is adopted to balance stability and efficiency. Stage one trains the phoneme/pitch encoders, various prediction networks, and the decoder. Stage two introduces the visual encoder and VCFM module, with the visual encoder frozen.

\subsection{Evaluation Metrics}

We use the following metrics for objective evaluation:
Mel Cepstral Distortion (MCD), F0 Frame Error (FFE), Singer Cosine Similarity (Cos) for audio quality, pitch accuracy and identity similarity,
and Lip‑Sync Error (LSE) for audio–video synchronization.
Specifically, Cos is computed from WavLM speaker embeddings fine-tuned for speaker verification~\cite{chen2022wavlm}, while LSE is computed based on audio and visual embeddings from the pre-trained SyncNet \cite{chung2016out}, with with LSE‑C (confidence) and LSE‑D (distance) as indicators.

For subjective evaluation, we use the Mean Opinion Score (MOS). For each setting, 20\% of test samples were rated on a 5-point scale by five evaluators—three with professional music knowledge and two music enthusiasts—across four dimensions: MOS‑Q (audio quality), MOS‑N (pitch naturalness), MOS‑S (timbre similarity), and MOS‑M (audio–video synchronization).

\section{Results and Analysis}
\subsection{Main Results}

Given that our model utilizes reference singing voice as input to restore timbre, we define four experimental settings: (1) same voice; (2) another voice of the same singer; (3) different singer; (4) unseen singer. Note that we do not use the audio content directly but compress it into a speaker embedding for timbre recovery, so the model targets the timbre of the reference voice rather than its content.
Since Setting 1 has a corresponding ground truth, we conduct both subjective and objective evaluations under this setting, reported in Table~\ref{tab:main}. For the remaining three settings, we conduct subjective evaluations. The overall MOS scores are computed and reported in Table~\ref{tab:subjective}.

As shown in Table~\ref{tab:main}, HPMDubbing-P yields only slight gains over the original model, and overall performance is limited. This indicates that even with pitch input and a frame-level predictor, HPMDubbing is still designed for movie dubbing and struggles with singing pitch dynamics. In contrast, StyleSinger performs substantially better, highlighting the effectiveness of its singing-specific pitch prediction and style modeling.

\begin{table}[t]
\centering
\small  
\setlength{\tabcolsep}{4pt} 
\caption{Subjective evaluation results (MOS) under setting 2-4.}
\label{tab:subjective}
\scalebox{0.95}{
\begin{tabular}{c|ccc}
\toprule
\textbf{Setting} & \textbf{Setting2} & \textbf{Setting3} & \textbf{Setting4} \\
\midrule
HPMDubbing        & 1.31 $\pm$ 0.42 & 1.25 $\pm$ 0.39 & 1.17 $\pm$ 0.33 \\
HPMDubbing-P & 1.51 $\pm$ 0.49 & 1.27 $\pm$ 0.49 & 1.19 $\pm$ 0.32 \\
StyleSinger       & \underline{2.69 $\pm$ 0.55} & \underline{2.30 $\pm$ 0.52} & \underline{1.88 $\pm$ 0.23} \\
\midrule
Our               & \textbf{2.75 $\pm$ 0.54} & \textbf{2.35 $\pm$ 0.61} & \textbf{1.98 $\pm$ 0.28} \\
\bottomrule
\end{tabular}}

\end{table}

Our method surpasses the second-best model, StyleSinger, achieving the highest scores across all objective metrics, with an MCD of 3.1125, FFE of 0.3921, COS of 0.9206, LSE-C of 1.4270, and LSE-D of 10.2782. These results indicate that our approach excels in restoring audio quality and pitch accuracy, enhances timbre fidelity, and improves audio–visual alignment. Furthermore, as shown in Table~\ref{tab:main} and Table~\ref{tab:subjective}, our method attains the highest MOS scores under every setting. Overall, these results demonstrate that our multimodal SVS framework effectively leverages visual cues, and that the visual supplementary fusion module efficiently incorporates visual information into audio synthesis, thereby improving both duration prediction and overall synthesis quality.

\subsection{Ablation Study}

We perform ablation studies on Setting 1 of the proposed dataset.

\begin{table}[t]
\centering
\setlength{\tabcolsep}{4pt}
\caption{Ablation study of fusion methods and training strategies. While -S and -T refer to the Single-Stage and Two-Stage training strategies, respectively. SMA denotes an attempted stepwise monotonic multi-head attention module, which was found to be unsuitable for this task.}
\label{tab:ablation}
\small
\scalebox{0.95}{
\begin{tabular}{l|ccccc}
\toprule
\textbf{Model} & \textbf{MCD$\downarrow$} & \textbf{FFE$\downarrow$} & \textbf{COS$\uparrow$} & \textbf{LSE-C$\uparrow$} & \textbf{LSE-D$\downarrow$} \\
\midrule
Baseline                   & 3.3580 & 0.4494 & 0.9141 & 1.3057 & 10.4129 \\
+ SMA                      & 3.9712 & 0.4538 & 0.9066 & 1.2653 & 10.8065 \\
+ VCFM-S      & 3.1307 & 0.4022 & 0.9195 & 1.3888 & 10.3146 \\
+ VCFM-T         & \textbf{3.1125} & \textbf{0.3921} & \textbf{0.9206} & \textbf{1.4270} & \textbf{10.2782} \\
\bottomrule
\end{tabular}}
\end{table}

\textbf{Effect of Fusion Methods.} As shown in Table~\ref{tab:ablation},We adopt StyleDubber's~\cite{cong2024styledubber} Stepwise Monotonic Multihead Attention~\cite{liang2020enhancing} for fusing textual and visual features.
However, this method yields suboptimal results across all metrics. we attribute this to singing's higher rhythmic and prosodic complexity, restricting certain fusion strategies. In contrast, the proposed VCFM module achieves significant improvements under single-stage training, demonstrating its effectiveness in fusing textual and visual features.

\textbf{Effect of Training Strategies.} As shown in Table~\ref{tab:ablation}, the two-stage training strategy  (first training the acoustic component, then incorporating visual features for joint optimization) outperforms single-stage training across all metrics, confirming the effectiveness of the staged approach.

\section{Conclusion}

In conclusion, we have constructed the first audio-visual SVS dataset and provided precise phoneme and pitch annotations. Additionally, we propose PerformSinger, a multimodal model that effectively integrates visual information for duration-free singing voice synthesis. We introduce an efficient fusion module that seamlessly incorporates visual features into textual representations, thereby enhancing
the model’s ability to leverage multimodal information. In future work, we will further explore more possibilities in multimodal SVS, including singing style transfer based on facial expressions and other related directions.

\bibliographystyle{IEEEbib}
\bibliography{strings,refs}

\end{document}